\documentclass[aps,prl,twocolumn,superscriptaddress]{revtex4}

\usepackage{graphicx}
\usepackage{dcolumn}
\usepackage{bm}
\usepackage{epsfig}
\usepackage{float}
\usepackage{subfigure}

\begin{document}

\title{Force distribution in granular media studied by an energy method based on statistical mechanics}

\author{S. A. Galindo-Torres}
\email{s.galinotorres@uq.edu.au}
\affiliation{School of Civil Engineering, The University of Queensland, Brisbane
QLD 4072, Australia}


\begin{abstract}
In the present letter a method to find a proper expression for the force distribution inside a granular sample in static equilibrium is proposed. The method  is based in statistical mechanics and the force distribution is obtained by studying how the potential elastic energy is divided among the different contacts between grains. It is found with DEM simulations with spheres that the elastic potential energy distribution follows a Bose Einstein law from which the force distribution is deduced. The present letter open a way in which granular materials can be studied with the tools provided by statistical mechanics.
\end{abstract}

\maketitle

The connection between the macroscopical behavior of granular materials and their microscopic interactions is still an open question with many unsolved problems. The main difficulty is that the methods proposed in statistical mechanics, the main branch of physics studying the micro-macro connection, cannot be easily used directly in granular materials due to their great energy dissipation. Considering this problems, Edwards and Bumenfeld\cite{blumenfeld2003granular} proposed a partition function based on the volume occupied by the grains and no on their individual energy. This certainly avoids the energy dissipation problems and also puts the volume of the ensemble as the main state variable of the granular ensemble. However the computation of the number of states associated with a given volume can be difficult because of many factors like for example the complex shape of particles. Since each particle has six degrees of freedom, the integral on the configuration space can be quite complex. Moreover the volumes of the particles interact with each other since the total volume must remain constant (as the energy in an ideal gas) which add even more complexity to the calculation of the configuration integrals. 

With DEM based simulations the force statistic have been studied and it has been found by Radjai, with Contact Dynamics simulations, that the distribution of the values of normal forces behaves like a power law before the mean value and like a exponential decay afterwards~\cite{radjai1996force}, however there is currently no justification for this particular behavior yet. Liu { \it et. al.}\cite{Liu1995} provided a new force distribution based on a mean field approximation called the {\it q}-model, but still it fails to properly represent small forces presented in the ensemble as discussed by Luding {\it et. al}\cite{luding2009force}.

In the present letter the author shows some results of force statistics obtained by DEM simulations with spheres in 3D. The force distribution is deduced by finding the distribution of potential energy in the contacts. The contacts are taken then as the new particles from which the ensemble is built. This particles, called by the author { \it Elastons}, represent the elastic interaction between two grains and as will be shown they greatly simplify the statistical mechanical study of granular materials.

\paragraph{The proposed model}
Although highly dissipative, granular materials follow the first law of thermodynamics as any other physical entity,
\begin{equation}
dU = \delta Q - \delta W,
\end{equation}
with $U$ the change in internal energy, $Q$ the heat dissipated by the system and $W$ the mechanical work done by the system. In granular materials this should be the template for every constitute model since the internal energy represents the potential energy stored by the elastic properties of the material, the heat represent the energy dissipated by frictional and viscous forces and hence the plastic deformation, and the work done relates the external stresses and strains. However given the difficulty in finding the functional form of each of these terms, the first law of thermodynamics is rarely the foundation of constitutive laws. In the present letter the author will focus his attention to the term related with the elastic domain: The internal energy $U$ that is not dissipated by viscous or frictional forces.

Once the kinetic energy of the grains is dissipated by internal viscosity and friction, the only mechanical energy remaining is the elastic potential energy stored in the contact points. This contact points are discrete and the total internal energy is just the addition of the individual potential energies at each contact point. These {\it Elastons} distribute the internal energy $U$ in the granular ensemble among themselves. Now in a granular ensemble of $N$ spherical grains there can be as much as $N(N-1)/2$ possible {\it Elastons} but most of them are not relevant since a spherical grain is only in contact with just a few other grains in general. However {\it Elastons} with non-zero energy are constantly being created and destroyed as the sample is subjected to external stresses representing the change in structure. 

In the initial state and in absence of external forces, the grains are just in contact with each other but there is no repulsion force. We called this state the {\it ground state} and all the {\it Elastons} have the same energy equal to zero. Now if stresses are applied to the sample some contacts start to accumulate potential energy and hence some {\it Elastons} jump from the ground state to higher energy levels. At the end of a compression test, most {\it Elastons} will have the lowest energy possible while some others will have higher energies. Considering that {\it Elastons} are in principle statistically indistinguishably, is tempting to assign a Bose-Einsten distribution form for their potential energy $E$,
\begin{equation}\label{eq:BES}
P(E) \propto \frac{g(E)}{\exp(\beta (E - \mu))-1},
\end{equation}
with $\beta$ related to the mean energy of the ensemble, $\mu$ a parameter that in the case of gases represent the chemical potential and $g(E)$ the density of states. Now, is dangerous to immediately  assign the same physical meaning of these quantities for the granular material as for the particle gases studied in statistical mechanics. For example the factor $\beta$ is related to the temperature of the gas as measured by a thermometer. But we cannot measure the $\beta$ factor in the same way as with the gases since in a gas the temperature is measured by radiation or convection, i. e. by the movement of particles. The {\it Elastons} do not move, their energy is entirely potential energy and hence the thermometer will not give us an accurate measurement of their {\it granular temperature}. At most, we can establish $\beta$ and $\mu$ as parameters to be found by fitting the above relation with experimental data.

But how to measure the potential energy of an {\it Elaston}? Clearly is very difficult to do so in an experimental setup. However the DEM method introduced by Cundall~\cite{cundall79} for the study of granular materials allows us to measure these microscopic quantities with ease. All that is needed is a contact law relating the potential energy and the force exerted by the grains.

In order to simplify the mathematical deductions the author has chosen the linear spring law that has been used recently for the simulation of the True Triaxial Test by Donze {\it et. al.}\cite{belheine2009numerical} instead of the more popular Hertz law \cite{luding97b}. In the linear spring law the spheres are assumed to overlap with each other and the elastic force is proportional to the overlapping length~\cite{galindo2009molecular}. The form of this force depends on a stiffness constant $K_n$ as,
\begin{equation}\label{eq:nforce}
\vec F = K_n \vec \delta,
\end{equation}
where $\vec \delta = \delta \hat n$ with $\hat n$ the normal vector joining the two sphere's centers and $\delta$ is the overlapping length.

With these simple model for the contact force, the energy of the corresponding {\it Elaston} can be found,
\begin{equation}\label{eq:potential}
E = \frac{1}{2} K_n \delta ^2
\end{equation}

In order to reach the static equilibrium state the whole kinetic energy has to be dissipated by viscous forces as used in previous studies~\cite{galindo2009molecular}. Once the static equilibrium state is reached, the internal energy is distributed among the {\it Elastons} as mentioned above. To apply Eq.~\ref{eq:BES} to the distribution of potential energy, the density of states for a given energy must be calculated. {\it Elastons} have only three degrees of freedom given by the components of the overlapping vector $\vec \delta$. The number of possible states $\Sigma (\delta)$ for an {\it Elaston} with overlapping length smaller than $\delta$ is then proportional to the volume integral in the configuration space,
\begin{equation}\label{eq:DOS}
\Sigma (\delta) \propto \int \int \int d\delta _x d\delta _y d\delta _z = \int 4 \pi \delta ^2 d\delta = \frac{4 \pi}{3} \delta ^3
\end{equation}
And hence the number of states with an energy smaller than $E$ is then by Eq.~\ref{eq:potential}:
\begin{equation}
\Sigma (E) \propto \frac{4 \pi}{3} \left ( \frac{2E}{K_n} \right )^{\frac{3}{2}}.
\end{equation}
Therefore the density of states can easily be found by differentiation of the previous function with respect to E,
\begin{equation}
g(E) = \frac{d\Sigma (E)}{dE} \propto \frac{4 \pi}{2} \left ( \frac{2}{K_n} \right )^{\frac{3}{2}} E^{\frac{1}{2}},
\end{equation}
and accordingly Eq.~\ref{eq:BES} is rewritten in the present case as:
\begin{equation}\label{eq:BESus}
P(E) \propto \frac{E^{\frac{1}{2}}}{\exp(\beta (E - \mu))-1}.
\end{equation}

Equations \ref{eq:BESus} and \ref{eq:nforce} can be used to obtain the distribution of forces $P(F)$ in the static equilibrium state of the granular ensemble. In order to do this, the reader must remember that for two random variables $x$ and $y$ related by a function $y=\Phi(x)$ their distributions follow $P(y) = P(x)\frac{d\Phi^{-1}(y)}{dy}$ and consequently,
\begin{equation}\label{eq:fdis}\label{eq:forcedis}
P(F) \propto \frac{F^2}{\exp \left (\beta \left (\frac{F^2}{2K_n} - \mu \right) \right)-1}.
\end{equation}

\paragraph{Simulations}
In order to check the distribution laws several True Triaxial Test (TTT) simulation have been carried out. In the TTT setup the spheres are enclosed by a set of six walls. Forces are applied to the walls in order to obtain a constant hydrostatic pressure over the sample. The simulation runs until the kinetic energy is dissipated. Once the static equilibrium state is reached the potential energy at each contact is measured and this is the energy associated to the {\it Elaston}. In the present letter some randomness is introduce by setting the spheres initially at a Hexagonal Close Packing and then randomly deleting some of them until the desired number of spheres is met. A total of 8 different hydrostatic pressures were considered with 8 random samples per case.

\begin{figure}[htb]
\centering
 \subfigure[]{\label{fig:force}\includegraphics[scale=0.38]{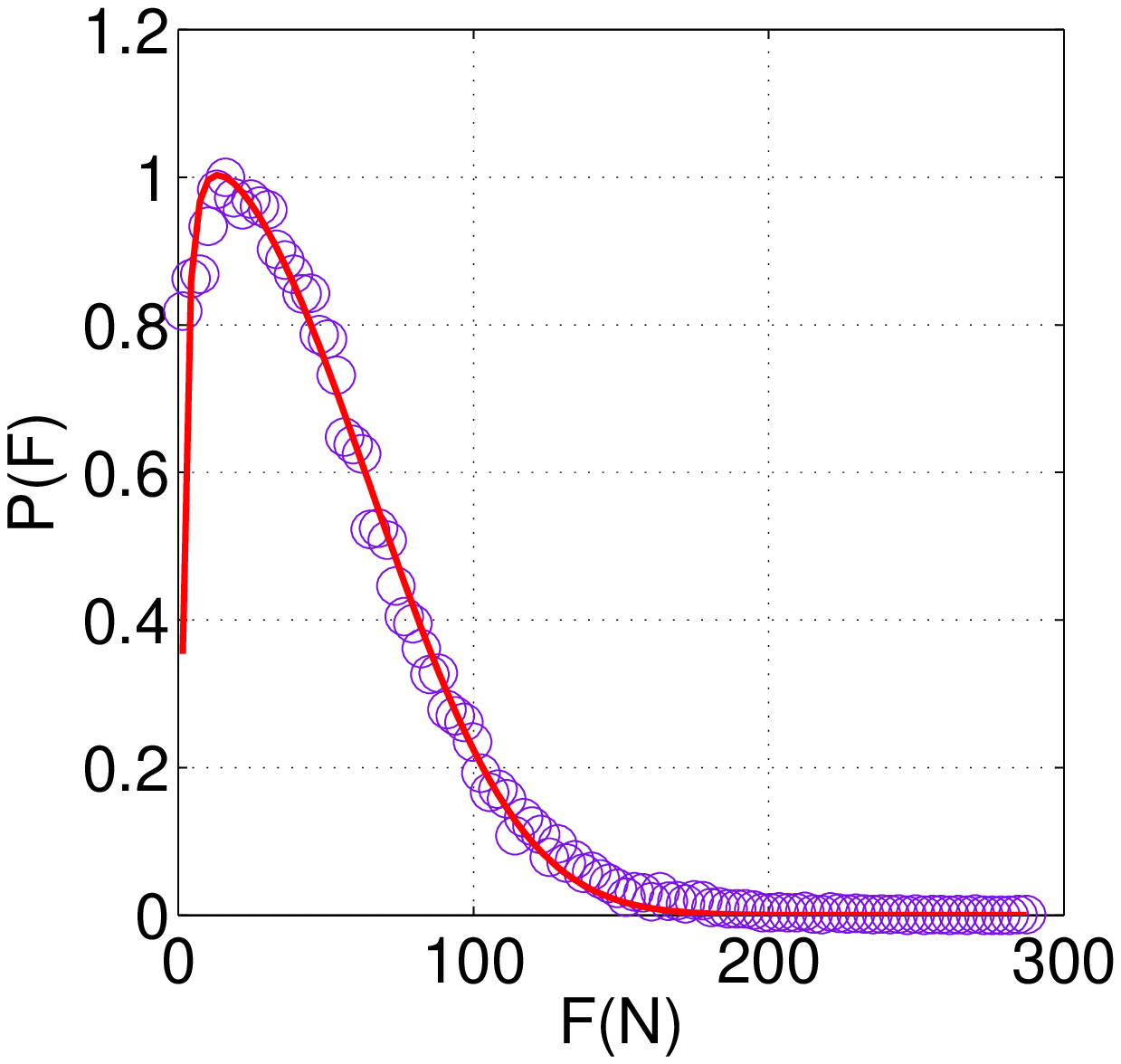}}
 \subfigure[]{\label{fig:forcelog}\includegraphics[scale=0.38]{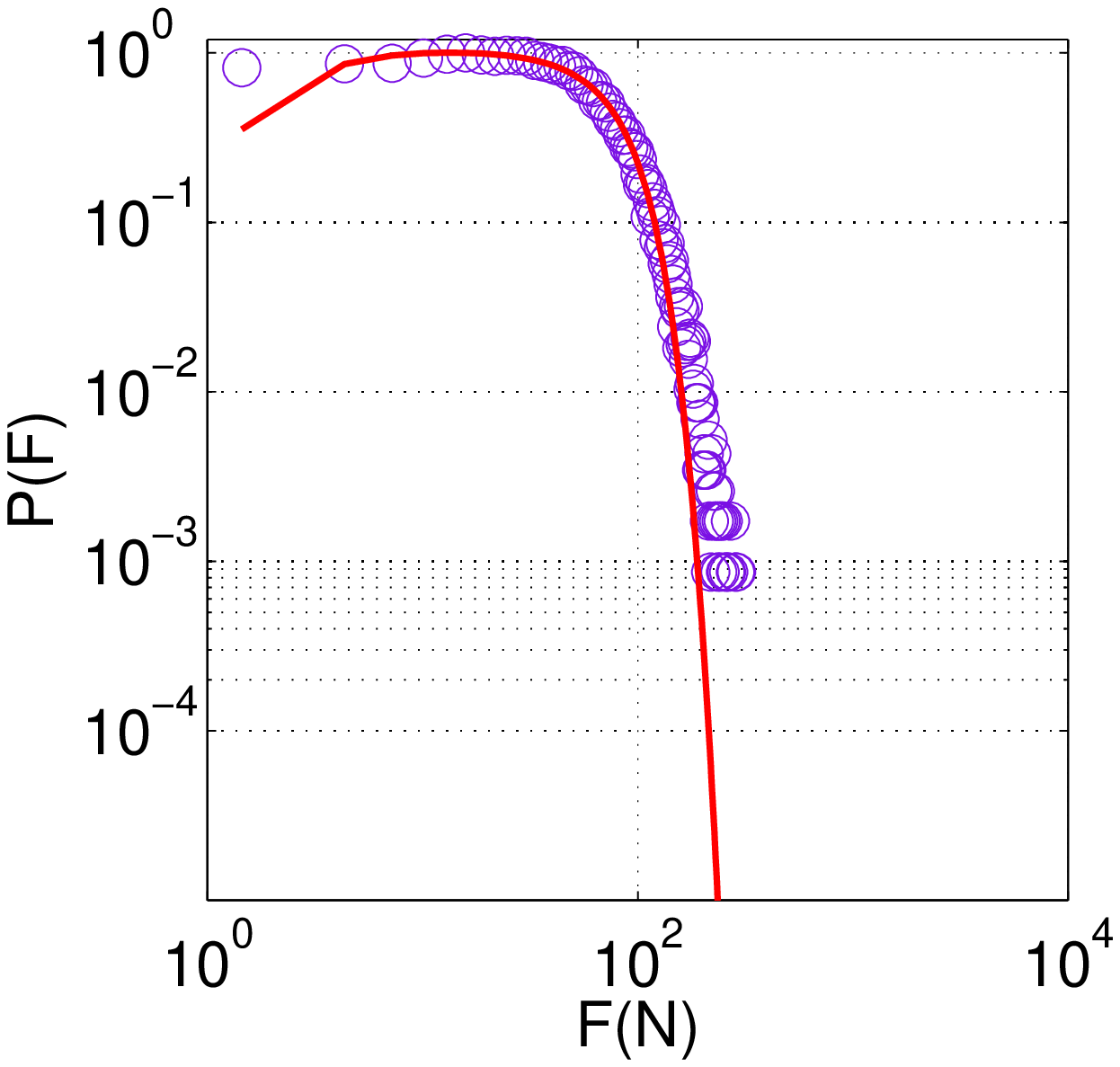}}
\caption{\label{fig:forcesta} Normalized force histogram (circles) for a given pressure in a) Linear-Linear scale, b) Linear Log scale and c) Log Log scale. The results are compared with the model of Eq.~\ref{eq:fdis} (solid line).}
\end{figure}

\paragraph{Results} For a given pressure the results for the force distribution at the static equilibrium test are shown in Fig.~\ref{fig:forcesta}. As can be seen the fitting is quite good with a $R^2$ factor of $0.999$ with values $\beta = 5.13 \pm 0.12 J^{-1}$ and $\mu = -0.0056 \pm 0.0009 J$. An important conclusion form the fitting process is that $\mu$ must be smaller than zero, otherwise there can be force values with negative population. In the simulation data the behavior observed by Radjai~\cite{radjai1996force} is also present as the force distribution can be approximated by a power law below certain value and by a exponential law after certain threshold. The model proposed in this letter does not consider this especial cases and covers the whole range of forces.

\begin{figure}[htb]
\centering
 \subfigure[]{\label{fig:energy}\includegraphics[scale=0.38]{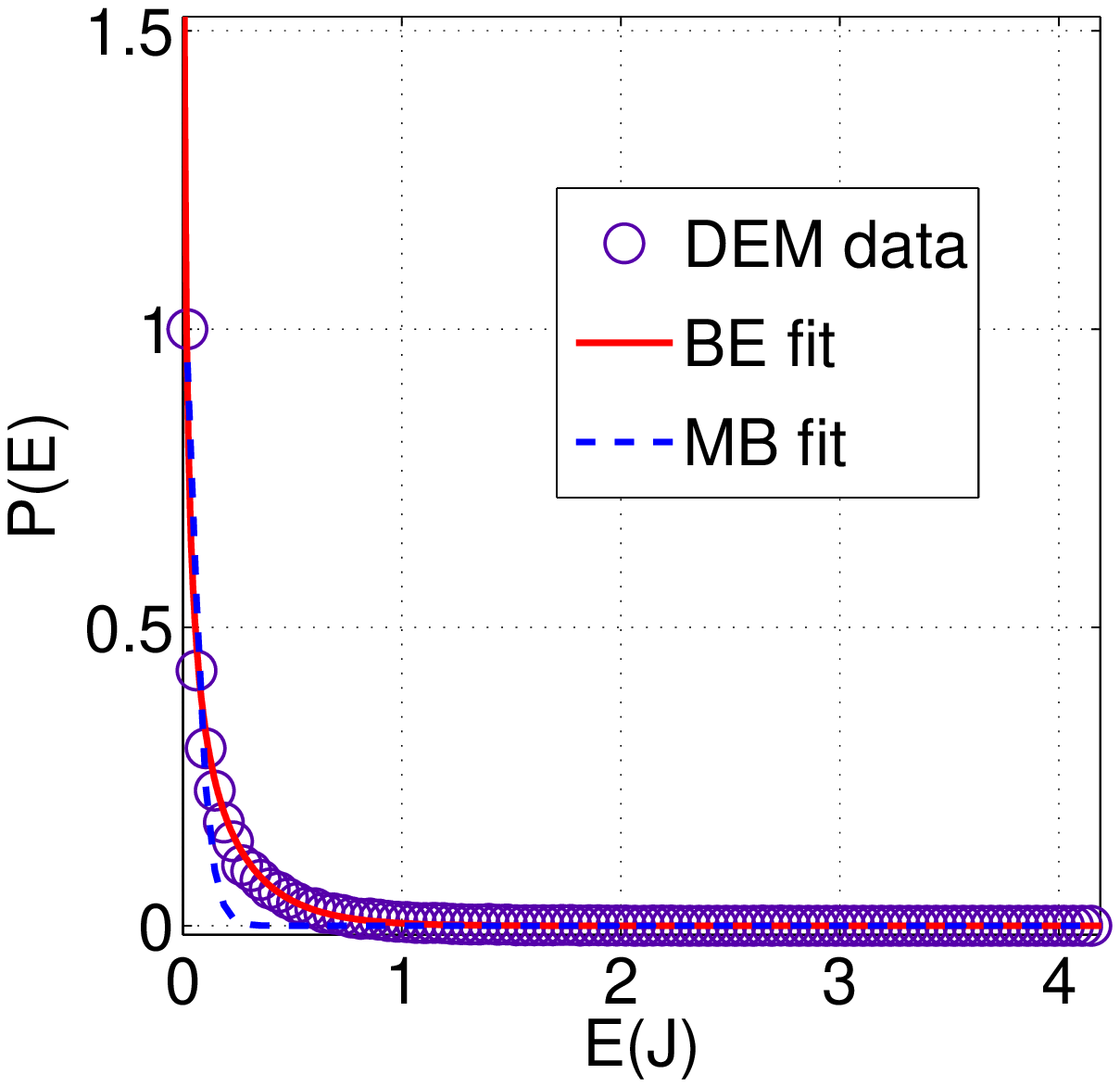}}
 \subfigure[]{\label{fig:energylog}\includegraphics[scale=0.38]{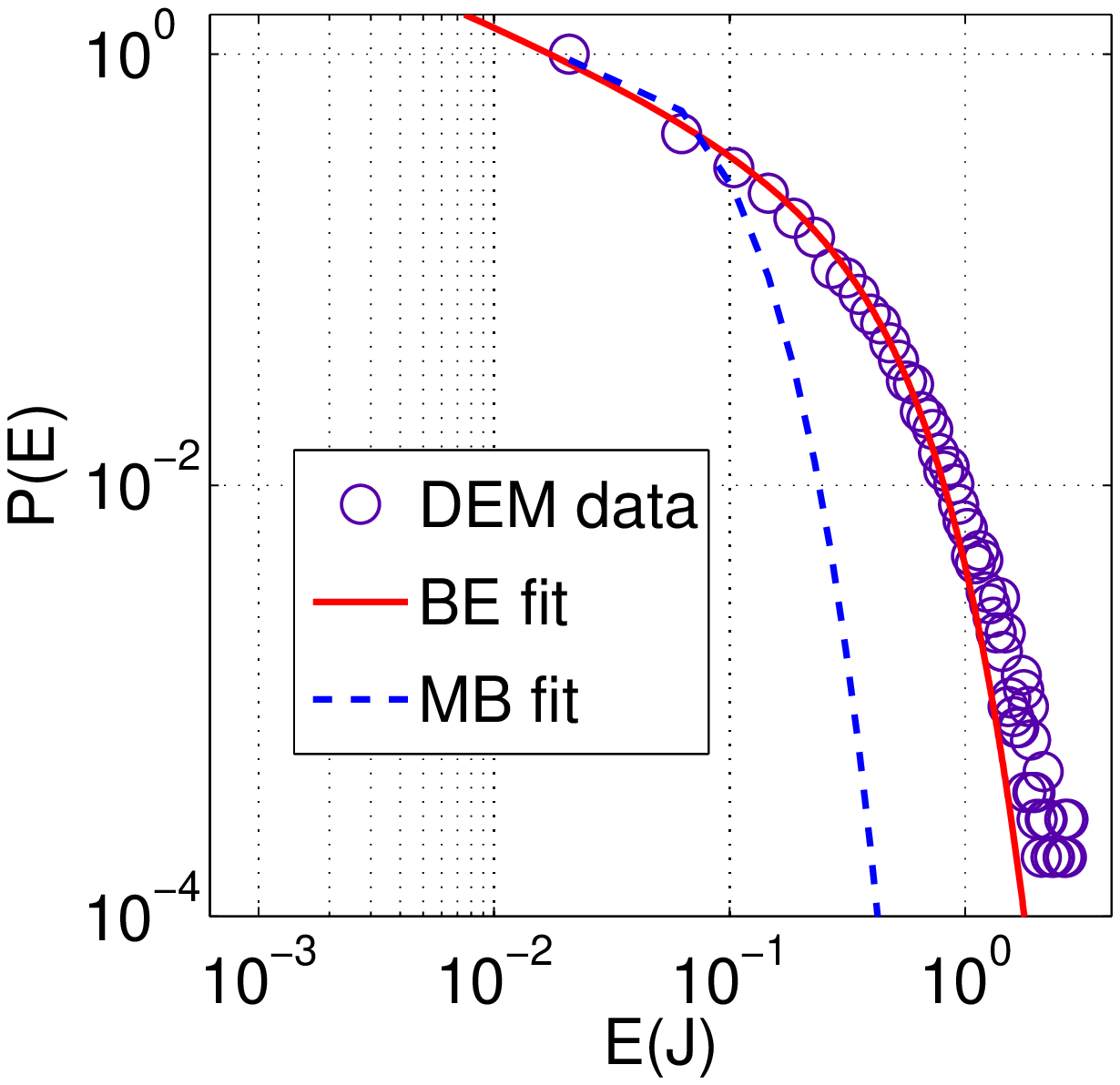}}
\caption{\label{fig:energyt} Normalized histogram of potential energy occupation (circles) for a given pressure in a) Linear-Linear scale, b) Log Log scale. The results are compared with the BE model of Eq.~\ref{eq:BESus} and also with the Maxwell Boltzmann (MB) model.}
\end{figure}

As the energy is the base of the proposed model, it was also fitted with Eq.~\ref{eq:BESus}. The results are shown in Fig.~\ref{fig:energyt} where the same values for $\beta$ and $\mu$ from the previous fitting have been used leaving only the proportionality constant as the only fitting parameter. Again the fitting has a good $R^2$ value of $0.999$. As expected the majority of {\it Elastons} lie in levels of small energies. It was also compared with the Maxwell Boltzmann statistics widely used for the macroscopic scale in which the population of energy states is given by:
\begin{equation}
P(E) \propto g(E) \exp \left ( -\beta E \right ) \propto E^{\frac{1}{2}} \exp \left ( -\beta E \right )
\end{equation}
Clearly by observing the DEM data, the BE model is a better fit than the classical MB model, confirming again that the {\it Elastons} behave statistically like Bosons.

\begin{figure}[htb]
\centering
 \label{fig:energyvsplog}\includegraphics[scale=0.38]{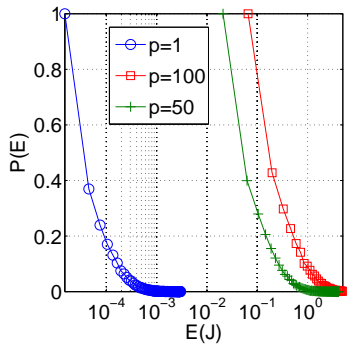}
\caption{\label{fig:energyvsp}Normalized energy histogram for three different pressures (expressed in Pa) in Log-Linear scale for detailed observation.}
\end{figure}

Next the dependence of the potential energy distribution on the applied hydrostatic pressure is observed. In Fig.~\ref{fig:energyvsp} the results for three different pressures are shown. The process of applying confining pressure to the Triaxial Test gives energy to the whole system and hence {\it Elastons} leave the ground state and occupy levels with higher energies. As expected the minimum pressure considered guaranties a low dispersion of {\it Elastons} over the energy spectrum, and the highest pressure ensures the largest dispersion.

\begin{figure}[htb]
\centering
 \subfigure[]{\label{fig:betavsw}\includegraphics[scale=0.38]{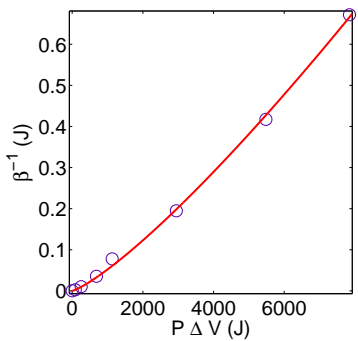}}
 \subfigure[]{\label{fig:betavse}\includegraphics[scale=0.38]{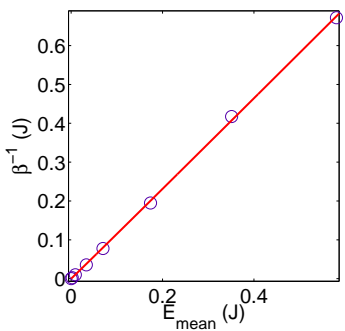}}
\caption{\label{fig:beta} Parameter $\beta$ versus a) the work done to the system $P \Delta V$ and b) The mean energy of the {\it Elastons}.}
\end{figure}

In a gas of particles, the dispersion of particles across the energy spectrum is usually related to the temperature. In the case of the present letter, the analogy should be done with the $\beta$ parameter of Eqs. \ref{eq:BESus} and \ref{eq:fdis}. Hence this parameter must be checked against the pressure applied to the confinement. This is done by plotting the inverse of $\beta$ against the work done on the sample expressed as $W = P \Delta V$ where $\Delta V$ is the difference in volume between the first pressure considered $p=1$ to the current pressure. In Fig.~\ref{fig:betavsw} it is shown how the inverse of $\beta$ grows with the external work done to the system by the pressure. It is not a linear relation, and its functional form is difficult to establish at the moment since the work done is divided in the potential elastic energy among the {\it Elastons} and in the dissipated energy that is not being considered here. 

The mean energy per {\it Elaston} is in contrast proportional to the inverse of $\beta$ (Fig.~\ref{fig:betavse}) as the mean energy of particles is proportional to the temperature in a gas. This is indeed expected an open a possible way to measure the mean potential energy by a special {\it granular temperature}.

\paragraph{Conclusions} In the present letter a method to study the statistics of force chains inside granular materials is proposed. the method is based in energy considerations. The question of how the potential elastic energy is distributed among the contacts is solved by a Bose Einstein (BE) law rather than a Maxwell Boltzmann (MB) one and, thanks to this, a force distribution is obtained and checked with DEM data. This implies that the contacts can be represented by particles that the author call {\it Elastons} and the tools provided by statistical mechanics for the study of Bosons gases can be use for the granular gas. The {\it Elastons} are far better suited for study since they have only three degrees of freedom in contrast with the six degrees that a grain in 3D space has. Moreover this particles do not interact with each other, which greatly simplifies the calculation of statistical properties. Future research in this area should focus in how the parameters of the proposed model area affected by the external stresses of the granular material, and that will help to the formulation of constitutive laws based on first principles.

\bibliographystyle{unsrt}
\bibliography{demref,dmpbib}
\end{document}